# Light transmission through and its complete stoppage in an ultra slow wave optical medium

**V. Ranjith[1]** and **N. Kumar[2*]**
[1]Centre for Quantum Information and Quantum Computation,
Department of Physics, Indian Institute of Science, Bangalore 560012, India.
[2]Raman Research Institute, Bangalore 560080, India.
*For correspondence: (e-mail: nkumar@rri.res.in)

**Abstract:** Light Wave transmission – its compression, amplification, and the optical energy storage – in an Ultra Slow Wave Medium (USWM) is studied analytically. Our phenomenological treatment is based entirely on the continuity equation for the optical energy flux, and the well known distribution-product property of Dirac delta-function. The results so obtained provide a clear understanding of some recent experiments on light transmission and its complete stoppage in an USWM.

Ultra slowing down of light and its complete stoppage in certain highly dispersive optical media is now an established experimental fact [1-5]. The extreme degree of dispersion required for this had been realized by use of *Electro Magnetically Induced Transparency* (**EIT**) [6]. In the present work we have studied analytically the case for a light wave traversing such an *Ultra Slow Wave Medium* (**USWM**). Based on the continuity equation for the optical energy flux, we derive a number of results such as wave compression, intensity growth (*i.e.,* amplification), and the wave energy storage peaking at the point of accumulation in the dispersive medium where the group velocity is at a minimum, even tending to zero, *i.e.,* complete stoppage. While these analytical results are interesting in their own right, they also seem to provide a clear phenomenological understanding of the experimental observations [2-5] of light wave stoppage and the associated energy storage, especially for the EIT – slowed light.

Consider the case of a light wave propagating through a 1D USWM of length *L* along x-axis $(0 \leq x \leq L)$. The strongly dispersive USWM is modeled here by a spatial profile for the group velocity $v_g(x) = c_0 \eta(x)$, where the slowness function $\eta(x) \to 1$ as $x \to 0$, and $\eta(x) \to 0$ as $x \to \Delta < L$ (see Fig. 1). Here $c_0$ is the speed of light outside the USWM. Thus, $x = \Delta$ is the point of accumulation for the wave energy in the USWM. (For a tunable USWM, the slowness parameter $\eta(x)$ can be switched on $(\eta(x)=1)$, or off $(\eta(x)=0)$ in time, and can thus allow even gating of the light wave.)

The energy flow through the USWM obeys the continuity equation

$$\frac{\partial I(x,t)}{\partial t} = -\frac{\partial}{\partial x}\left(v_g(x) I(x,t)\right) - \frac{I(x,t)}{\tau} \quad 0 \leq x \leq \Delta \quad , \tag{1}$$

where $1/\tau$ is the energy dissipation rate (as the medium may be lossy) giving an exponential attenuation of the wave intensity $I(x,t)$. (Note that in the case of non-zero loss, the group velocity $v_g$ should strictly speaking be replaced by the energy flow velocity $v_E$ [7]. We will, however, ignore this finer point here, and continue to use the group velocity $v_g$ for the purpose of our phenomenology).

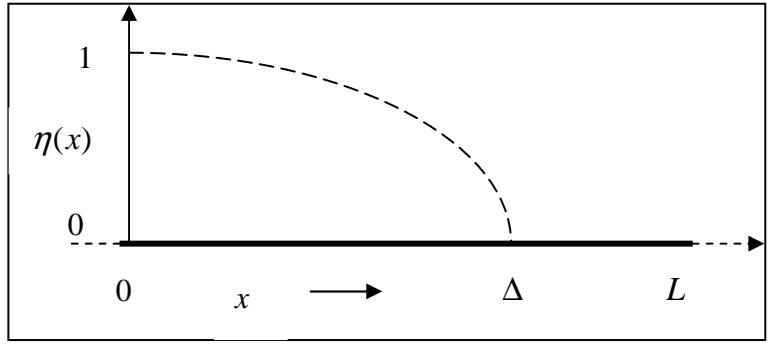

Fig. 1. Group velocity profile $\eta(x)$ for the model USWM $(0 \leq x \leq L)$, with the point of accumulation $\Delta$, where the group velocity vanishes (Schematic).

Now, consider the light wave of intensity $I(0)$ from a cw laser, say, to be incident on the USWM at $x = 0$. As the light cannot escape beyond the point of accumulation $x = \Delta$, we expect a steady state to reach when the incident energy flux is balanced by the dissipation in the USWM because of the loss rate $(1/\tau)$. The steady-state energy stored in the USWM can be readily calculated. Let $I_{ss}(x)$ be the steady-state intensity in the USWM. The total wave energy stored in the steady state $U_{ss}$ is obtained by equating the rate of the influx $c_0 I(0)$ with the loss rate $U_{ss}/\tau$, giving

$$U_{ss} = c_0 I(0) \tau \qquad (2a)$$

This is, of course, independent of the detailed shape of $\eta(x)$ for the USWM.

Next we consider the intensity profile $I_{ss}(x)$ in the steady state. Setting $\frac{\partial I}{\partial t}(x,t) = 0$ in Eq. (1), we get the equation for the steady-state intensity

$$0 = -\frac{\partial}{\partial x}\left(c_0 \eta(x) I_{ss}(x)\right) - \frac{I_{ss}(x)}{\tau} \qquad 0 \leq x \leq \Delta \;, \qquad (2b)$$

giving,

$$\frac{\partial I_{ss}(x)}{\partial x} = -\frac{1}{c_0 \eta(x)}\left(c_0 \frac{\partial \eta(x)}{\partial x} + \frac{1}{\tau_0}\right) I_{ss}(x) + A\delta\left(1 - \frac{x}{\Delta}\right) \ldots \qquad (2c)$$

The last term $A\delta\left(1-\dfrac{x}{\Delta}\right)$ in Eq. (2c) arises because of the division by $\eta(x)$ which vanishes at the accumulation point $x=\Delta$. (Such a term follows from the Dirac distribution-product identity, viz., for any $f(x)$ with $x\,f(x) = x\,g(x)$, we have $f(x) = g(x) + k\delta(x)$, where the constant $k$ must be determined from, e.g., a normalization condition. Recall that $\eta(x)\delta(x-\Delta) = 0$ for our model of the slowness function $\eta(x)$ as above, and the light wave is confined to $x\leq\Delta$.)

Equation (2c) is readily solved to give for $x<\Delta$,

$$I_{ss}(x<\Delta) = I(0)\,\frac{1}{\eta(x)}\,\exp\left(-\frac{1}{c_0\tau}\int_0^x \frac{1}{\eta(x')}\,dx'\right), \quad \text{for } x<\Delta. \qquad (3)$$

Thus, the point $x=\Delta$ will contribute a delta-function at $x=\Delta$ so as to give the total steady-state energy stored in the USWM equal to $U_{ss}$, consistent with Eq. (2a). For the purpose of explicit illustration now, the USWM will be modeled as $\eta(x) = \left(1 - x/\Delta\right)^\alpha$, with the slowness-exponent $0<\alpha<1$.

The condition $0<\alpha<1$ ensures a finite time $\displaystyle\int_0^\Delta \frac{dx}{c_0\eta(x)} = \frac{\Delta}{c_0(1-\alpha)}$ for the wave to reach the point of accumulation $\Delta$. (Of course, $\alpha>1$ is not ruled out, and can be treated in the same way.) With this choice, we have the steady-state solution

$$I_{ss}(x) = \frac{1}{(1-x/\Delta)^\alpha}\cdot\exp\left(-\frac{\Delta}{c_0\tau}\,\frac{1-(1-x/\Delta)^{1-\alpha}}{1-\alpha}\right) + \text{a boundary term}, \qquad (4)$$

where, the boundary term is localized at $x=\Delta$. It is a delta function that will be fixed now as follows.

The steady-state intensity $I_{ss}(x<\Delta)$ together with the delta-function term located at the boundary $x=\Delta$ must give the total integrated stored energy $\displaystyle\int_0^\Delta I_{ss}(x)\,dx = U_{ss}$ as obtained in Eq. (2a). This at once fixes the boundary term uniquely to be $2c_0\tau\exp-\left(\dfrac{\Delta}{c_0\tau(1-\alpha)}\right)\,\delta(x-\Delta)$.

With this, we finally have the complete steady-state solution

$$I_{ss}(x) = \frac{1}{(1-x/\Delta)^\alpha} \cdot \exp\left(-\frac{\Delta}{c_0\tau}\frac{1-(1-x/\Delta)^{1-\alpha}}{1-\alpha}\right) +$$

$$+ \quad 2c_0\tau \exp-\left(\frac{\Delta}{c_0\tau(1-\alpha)}\right) \delta(x-\Delta) \qquad (5)$$

This is the main result of our paper for the model. It is plotted in Fig. 2. It has in it the essential features of wave compression, wave amplification and the energy storage in the USWM as stated above. It clearly shows the competition between the loss ($1/\tau$) and the group velocity dispersion ($\eta$). Note the power-law divergence co-existing with the *delta-function condensation* (pile-up) of the wave intensity at the point of accumulation $x=\Delta$.

The intensity peaking (which in fact happens to be an integrable singularity for the above model choice of $\eta(x)$) is important to our phenomenological interpretation of any experiment on light storage in a USWM.

It is to be noted, however, that in the limit of $\tau \to \infty$ (that is for no dissipation) and with a finite incoming optical pulse, the light wave simply *piles up* to a complete stop at $x = \Delta$. This should be the case for a USWM as realized in a BEC [1], and will manifest as a *bright spot* appearing at the point of stoppage $x = \Delta$.

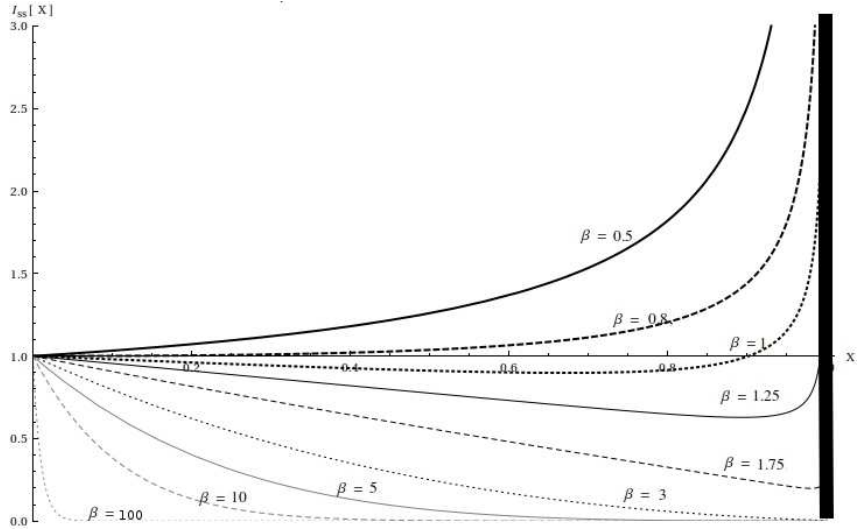

Fig. 2. Plot of steady-state intensity $I_{ss}(X)$ against the normalized distance $x/\Delta \equiv X$ in a USWM for the choice of parameters: $\alpha = 0.8$, and $\Delta/c_0\tau \equiv \beta$. Here, $\beta$ ranges from 0.5 to 100. Note the thick vertical line at $x/\Delta = 1$ representing the delta-function peaks at the point of accumulation (the bright spot).

It is important to emphasize here that the above phenomena of wave compression, intensity growth and the energy storage are generic to any USWM. One may note in passing that this phenomenon of wave compression and amplification is, of course, analogous to that of *breaking* of waves at the sea shore: as the wave train approaches the shore, the leading edge (being in shallower waters) advances slower relative to its trailing edge giving wave compression. This in turn leads to amplitude growth or a pile-up of the compressed wave, and then to its eventual *breaking* close to the shore line. The sea shore acts as the slow wave medium here.

The above analytical treatment is, of course, readily generalized to the case of an arbitrary group velocity profile other than the one shown in Fig. 1. This may include, in particular, the case where the point of accumulation (complete stoppage) gets replaced by a broad maximum, allowing the transmission of light beyond the peak.

In conclusion, we have proposed and solved analytically a model for light propagation in an Ultra Slow Wave Optical Medium having a point of accumulation where the group velocity becomes extremely low, or even vanishes. Our solution explicitly displays wave compression (pile-up), optical energy storage and a point of accumulation that characterize the USWM.

**Acknowledgments:**
One of us (RV) thanks Prof. Anil Kumar and acknowledges DST-0955 project (CQIQC) at IISc for support, and also thanks RRI for its hospitality during the course this work.